\documentclass[preprint,12pt]{elsarticle}

\usepackage{amssymb}
\usepackage{url}
\usepackage{array}
\usepackage{xcolor}
\usepackage{siunitx}

\graphicspath{{figs/}}

\journal{Applied Materials Today}

\begin{document}

\begin{frontmatter}

\title{Learning in colloids: Synapse-like ZnO + DMSO colloid}

\author[uwe]{Noushin Raeisi Kheirabadi*}
\ead{Noushin.Raeisikheirabadi@uwe.ac.uk}
\author[iit,uwe]{Alessandro Chiolerio}
\author[uwe]{Neil Phillips}
\author[uwe]{Andrew Adamatzky}

% \affiliation[uwe]{organization={Unconventional Computing Laboratory, UWE Bristol},%Department and Organization
%            postcode={BS16 1Y}, 
%            country={UK}}

\affiliation[uwe]{organization={Unconventional Computing Laboratory,  University of the West of England, Bristol, UK}}

\affiliation[iit]{organization={Center for Bioinspired Soft Robotics, Istituto Italiano di Tecnologia,  Genova, Italy}}

\begin{abstract}
Colloids submitted to electrical stimuli exhibit a reconfiguration that could be used to store information and, potentially compute. We investigated learning, memorization, and time and stimulation's voltage dependence of conductive network formation in a colloidal suspension of ZnO nanoparticles in DMSO. Relations between critical resistance and stimulation time were reconstructed. The critical voltage, i.e. the stimulation voltage necessary for dropping the resistance, was shown to decrease in response to an increase in stimulation time. We characterized a dispersion of conductive ZnO nanoparticles in the DMSO polymeric matrix using FESEM and UV–visible absorption spectrum.
\end{abstract}

%%Graphical abstract
%\begin{graphicalabstract}
%\includegraphics{grabs}
%\end{graphicalabstract}

%%Research highlights
%\begin{highlights}
%\item Research highlight 1
%\item Research highlight 2
%\end{highlights}

\begin{keyword}
Neuromorphic device \sep Colloid \sep Memory devices \sep Artificial synapse \sep liquid computing \sep liquid robotics
\end{keyword}

\end{frontmatter}

\section{Introduction}
\label{sec:sample1}

Massive parallel computations could be implemented using colloids since they have a wide range of unique properties.  They are stable, meaning the particles remain suspended in the solution. Therefore, any particle can be addressed to access a memory cell in a colloid processor \cite{adamatzky2021handbook}. Moreover, colloid computers could be fault tolerant due to redundancy of the elementary processors-particles. 

A most suitable candidate for colloid-based computers would be a colloid of conductive particle in insulating solution. In this paper we use zinc oxide particles suspended in a solution of polymer dimethyl sulfoxide.
Polymers in their natural state (with some relevant exceptions) are insulating materials~\cite{doi:10.1021/acs.chemrev.9b00766}, and the modulation of their electrical conductivity, regardless of the application, requires blending with materials featuring electronic / ionic conductivity. For example fillers in the metallic state allow increased electronic conduction, and depending on their shape and aspect ratio, can create a complete network providing a preferential route for charge carriers. An example is given by silver nanoparticles / nanowires ~\cite{mme2013} that are frequently used as conductive fillers. The metallic state can be achieved to some extent also by carbon-based materials, allowing to keep a good balance towards the raw materials supply. Some examples: graphene, carbon fibres, carbon nanotubes that are compounded into a common polymer over a threshold concentration~\cite{sumita1986electrical}. Such threshold is known as the critical concentration, above which percolation occurs (U)~\cite{wessling1991electrical, sumita1986effect, stauffer1979scaling}. Percolation appears after a phase transition, where a dramatic change occurs at a single sharply defined parametric value as the concentration is modified continuously~\cite{zhang2006temperature, zhang2005electrical}. Lowering a composite system's percolation threshold looks to be an effective strategy to reduce the quantity of (precious) filler necessary to achieve appropriate conductivity and, as a result, avoid mechanical performance issues~\cite{grunlan2001lowering}.
Attempts to employ alternative percolation models to describe conductive polymer composites have demonstrated that no single model can explain all of the different experimental results~\cite{castellino}. This discrepancy could be related to the fact that conductivity is highly dependent on the geometric parameters of the filler, the quality of their bonds, and the matrix-filler interaction~\cite{miyasaka1982electrical,sumita1986effect}. Until now, most research on the dynamic process of conductive network generation under an electric field has concentrated on composites containing conductive fillers like carbon black (CB), carbon fibres (CF), or carbon nanotubes (CNT)~\cite{liu2016role, li2012thermal}. Nevertheless, and we reach now a fundamental point about our research, there is no straightforward way to change conduction properties, once a composite material is formed and a physical form has been given. Having an easily tuneable material would enable us to create switches, memories and other sorts of devices fundamental for computing. Chemical reactions involving different oxidation states might also be exploited, triggered by high electric fields, nevertheless activation energies should be overcome and peculiar fabrication conditions be met, so that from the practical point of view, applications are not so easy ~\cite{porro2017, RSCA2016}.

Nanofluids, or colloidal dispersion of nanoparticles in a liquid solvent, have received a lot of attention in recent years because of their prospective applications~\cite{zhang2008stability}, encompassing intrinsic plasticity, self-adaptability and fault-tolerance~\cite{chiolerio2017smart, chiolerio2020liquid}.
Even after long periods of storage, a stable colloid is expected to remain sediment-free. The size and density of the dispersed particles determine the settling behaviour of dispersions~\cite{anand2017role}.
The liquid dispersion of nanopowders is a challenging process. The high surface area and surface energy of nanomaterials, which are responsible for their favourable effects, generate particle agglomeration, resulting in poor-quality dispersions.

Research on quantum-sized semiconductor particles has exploded in the last ten years, due to their fascinating novel optical and electrical capabilities~\cite{cohen2000theory}. ZnO is a semiconductor featuring a wide band gap and a large exciton binding energy of 60 meV at room temperature, being a potential choice for room-temperature UV lasers or other fascinating excited-state phenomena~\cite{dong2005preparation, acsami2015}. ZnO is a versatile material with high electrical conductivity, visual transparency, photostability, and photocatalytic abilities~\cite{lee2020high,marin2009thermal}.

The percolation threshold for simple binary composites depends entirely on the aspect ratio of the filler, and ranges between approximately 10 \% up to 70 \%, as anticipated by classical percolation theory for a random system~\cite{tang1996studies, scher1970critical}.  After a percolation threshold is achieved, random resistor networks usually follow a power-law conductivity relationship:
$$
\sigma = \sigma_0 (V-V_c)^s
$$
where $\sigma$ is the composite conductivity (S/cm), $\sigma_0$ is the intrinsic conductivity of the filler, s is the power-law exponent (typically 1.6–2.0 in 3D), and V$_c$ is the volume fraction of filler at the percolation threshold (near 0.15 for random fiber-based 3D systems)~\cite{kirkpatrick1973percolation}.

 Von Neumann computer architecture (in which memory and the CPU are separated) only allows instructions to be carried out one at a time and sequentially, which restricts parallel operations~\cite{akhter2006multi}. The next generation of computing systems are anticipated to use synthetic synapses (which mimicking biological synapses) enabling information to be processed in parallel~\cite{han2019recent}. In principle, resistive switching and synaptic properties of learning colloids might be utilised to enable massively parallel computing~\cite{van2018organic}.

The aim of the present study is to construct highly mobile ion-conductive pathways using ZnO nanoparticles in the DMSO solution. The free migration of ions is expected to provide high ionic conductivity. We study how, having a spherical symmetry filler and a 3D composite fluid, a tuneable percolation threshold can be achieved with low concentrations.

\section{Methods}

ZnO nanoparticles and DMSO were purchased from US research nanomaterials, Sodium Dodecyl Sulphate (SDS) and Sodium Hydroxide (NaOH) were purchased from Merck. De-Ionised Water (DIW) were prepared in the lab with Millipore de-ionised water generator device, model Essential, rated 15 Mohm cm. 

Surfactant solution with a concentration of 0.22 wt\% was prepared by adding SDS in DIW and stirred to get a homogeneous solution. 1 mg ZnO nanoparticles were added to DMSO with continuous stirring. Then 2 ml of SDS solution and 1 ml NaOH 10M added to the mixture under stirring. The concentration of resulting dispersion was maintained at 0.11 mg/ml. The resulted suspension was placed in an ultrasonic bath for 30 minutes. Then stirring process was continued for a few more hours to get a uniform dispersion of ZnO~\cite{anand2017role}.

All electrical measurements were made with Fluke 88464A and Keithley 2400. Field emission scanning Electron Microscopy (FEI Quanta 650 FESEM) was also used to characterize the nanoparticle suspensions. The accelerating voltage was set to 10 kV in this study, and the working distance was set to around 5 mm. The images' contrast and brightness were optimised so that particles could be easily distinguished from the background.
Ultraviolet-visible (Uv-Vis.) spectrometer (Perkin Elmer Lambda XLS) was used to measure sample absorbance at room temperature.

\section{Results and Discussions}

\subsection{Suspension structure}

\begin{figure}[!tbp]
\includegraphics[width=\textwidth]{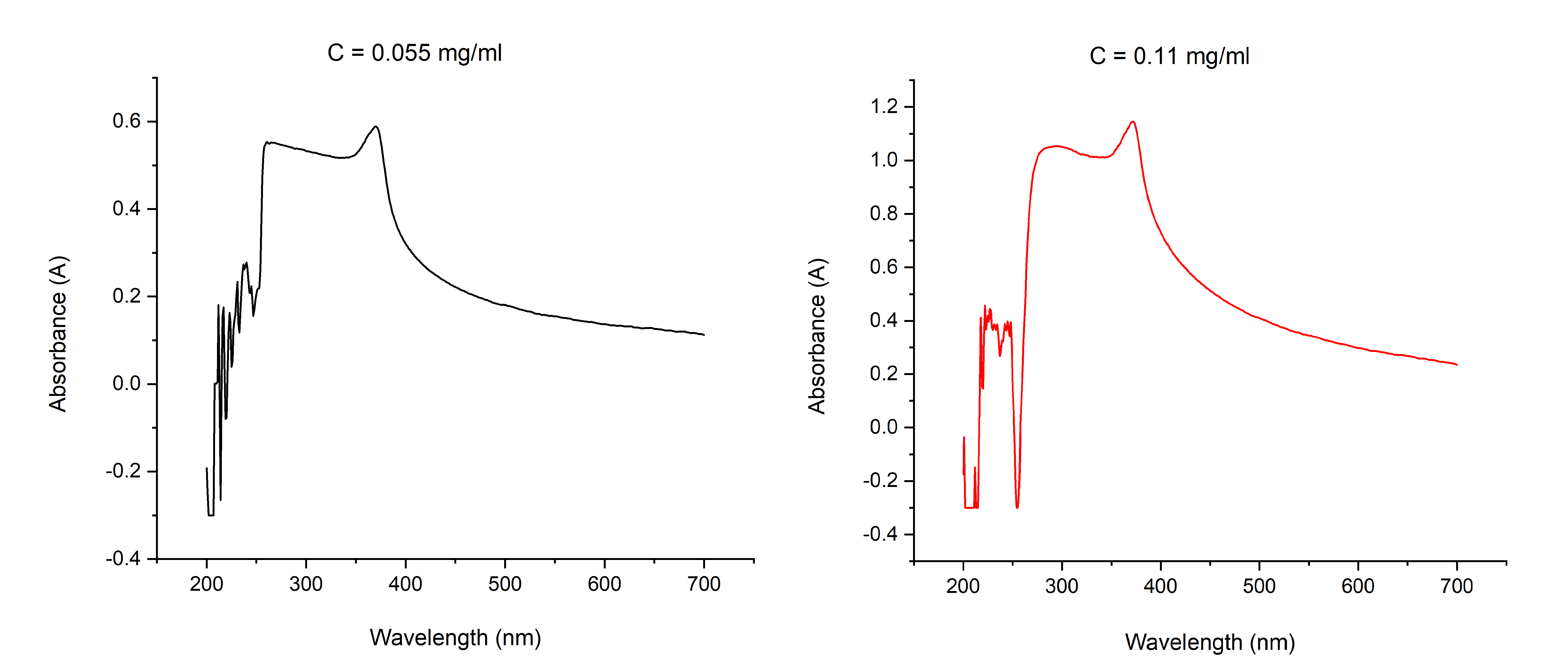}
\caption {UV-Visible spectra of the sample with concentrations 0.055 and 0.11 mg/ml}
\label{fig01}
\end{figure}

\begin{figure}[!tbp]
\includegraphics[width=\textwidth]{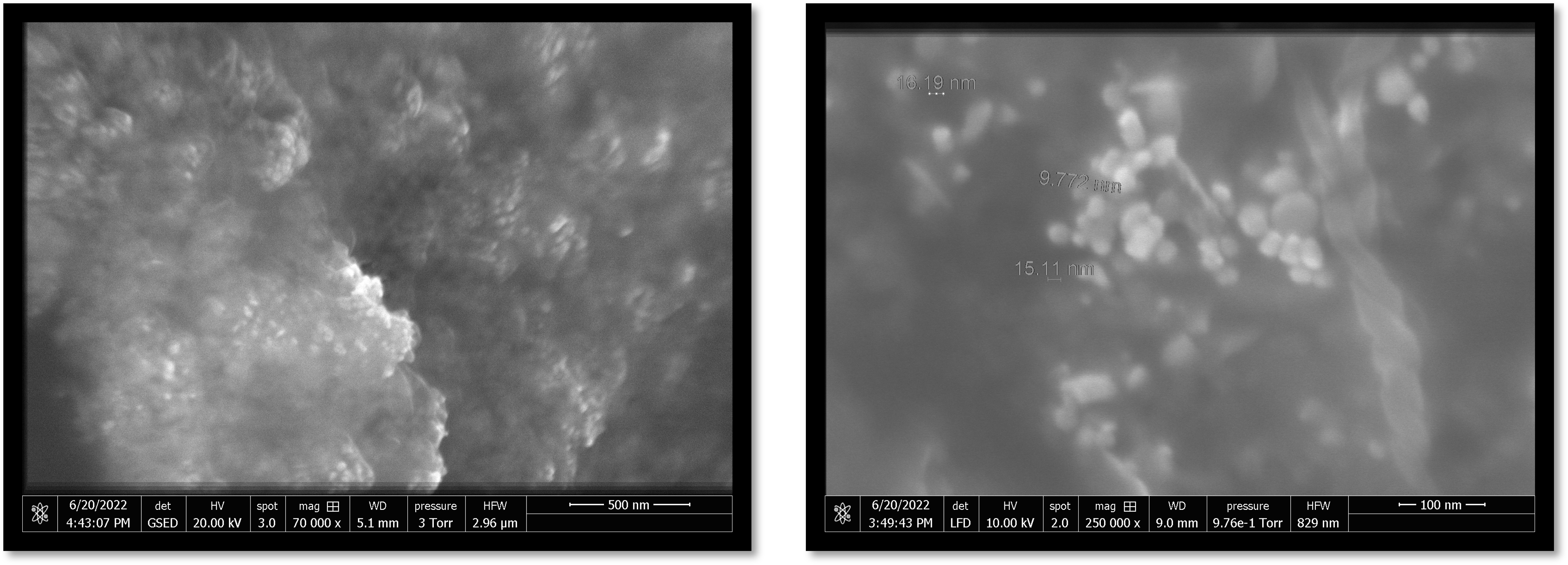}
\caption {SEM images of drop casted ZnO colloids on Copper substrate in two different magnifications}
\label{fig02}
\end{figure}

The UV–visible absorption spectrum of ZnO colloids, with different concentration, at room temperature is recorded in the wavelength range of 200–700 nm. Fig.~\ref{fig01} depicted resulted plots of UV–visible absorption spectrum. The spectra have peaks at 370 nm (colloid with concentration of 0.055 mg/ml) and 372 nm (colloid with concentration of 0.11 mg/ml). These absorption  peaks at 370-372 nm are the characteristic peak for hexagonal ZnO nanoparticles~\cite{pudukudy2015facile}. In comparison to bulk ZnO (370 nm), there is a good agreement with previous reports~\cite{reddy2011combustion,sun2011enhanced}. Calculation of the optical band gap was based on the below equation:
$$
E_g (eV) = hc/\lambda = 1240/\lambda
$$
Where $E_g$ is the optical band gap, h is the Planck’s constant, c is the speed of light,and $\lambda$ is the wavelength of maximum absorption. Here Optical band gap was calculated as 3.35 eV which matches well with previous reports~\cite{reddy2011combustion,baskoutas2010conventional,baskoutas2011transition}.

In order to study particles morphology and size by FESEM technique, a thin layer of ZnO colloid were prepared by drop casting a drop of ZnO particle suspension (0.11 mg/ml) on a Copper foil with a thickness of 100 $\mu$m at room temperature.

FESEM results in Fig.~\ref{fig02} depict particle agglomeration during sample preparation.As a result of the surface tension of the solvent during evaporation, FESEM observation rarely reveals separated spheres, and most of the ZnO spheres are multilayered. Indeed, increased liquid surface tension would draw the nanoparticles closer together, causing them to re-aggregate during the drying process~\cite{lu2018methodology}.

\subsection{Suspension Conductivity}

To study a formation of conductive network of nanoparticles we carried out  DC stimulation of the suspension mixture with different parameters of the stimulation. 

\begin{figure}[!tbp]
\centering
\includegraphics[width=\textwidth]{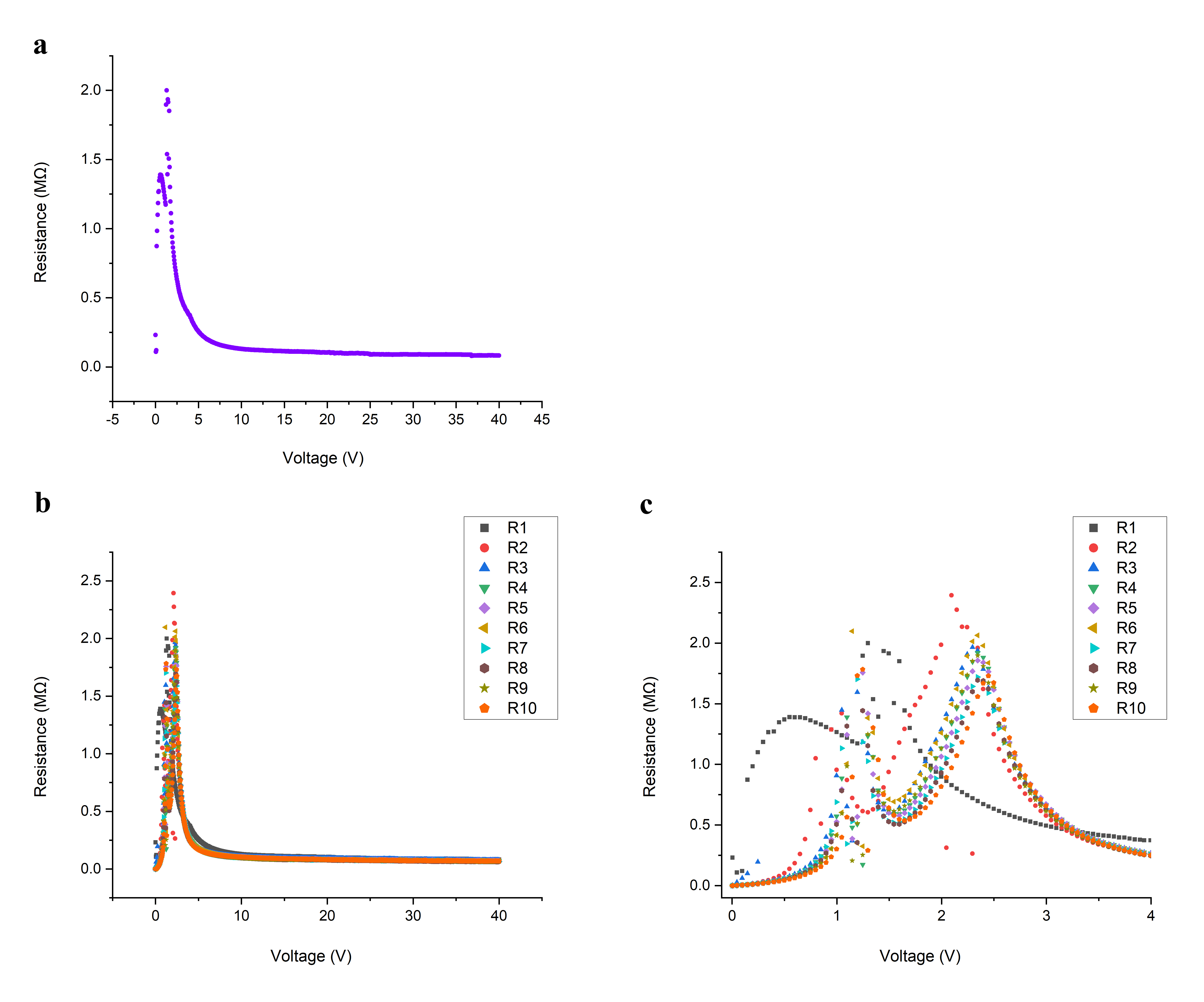}
\caption {a) Resistance diagram of colloid after 40V DC stimulation, b) Repeated measurement resistance of colloid with stimulation till 40V, c) plot section b with zoom on the first area of the curve}
\label{fig03}
\end{figure}

In the first step, DC stimulation versus time at the rate of 10 mV/s from \SIrange{0}{40} {\volt}, with Keithley, have been applied to the sample and current was measured. Results are shown in Figs.~(\ref{fig03},a). The increase in voltage, resulting in a decrease in resistance. So the first step experiment is: 1) apply voltage growing from 0 to 40, 2) measure resistance, 3) Repeat.

To check the repeatability of this result, DC stimulation has been applied from \SIrange{0}{40} {\volt}, with Keithley, to the sample for 10 times and then current was measured. All data are integrated in the Fig.~\ref{fig03}(b,c). Figure (3b) shows the repeated measurement resistance of sample with stimulation till 40~V and figure (3c) shows the same plot with zoom on the first area of the curve. Just the first stimulation has a little different resistance curve in comparison with the other 9 stimulations, as it often occurs in similar systems such as resistive switching devices. An interesting feature is the presence of maximum resistance, found in the range of 2 to 2.5~V and then it drops down. As bubbling was observed during apply voltage, probably from 2~V till 3~V, bubbling accrues and misses the order of nanoparticles so R increases. Perhaps after 3~V, the voltage is enough strong to overcome bubbling turbulence and reorder nanoparticles again so R decreases again.

\begin{figure}[!tbp]
\centering
\includegraphics[width=0.8\textwidth]{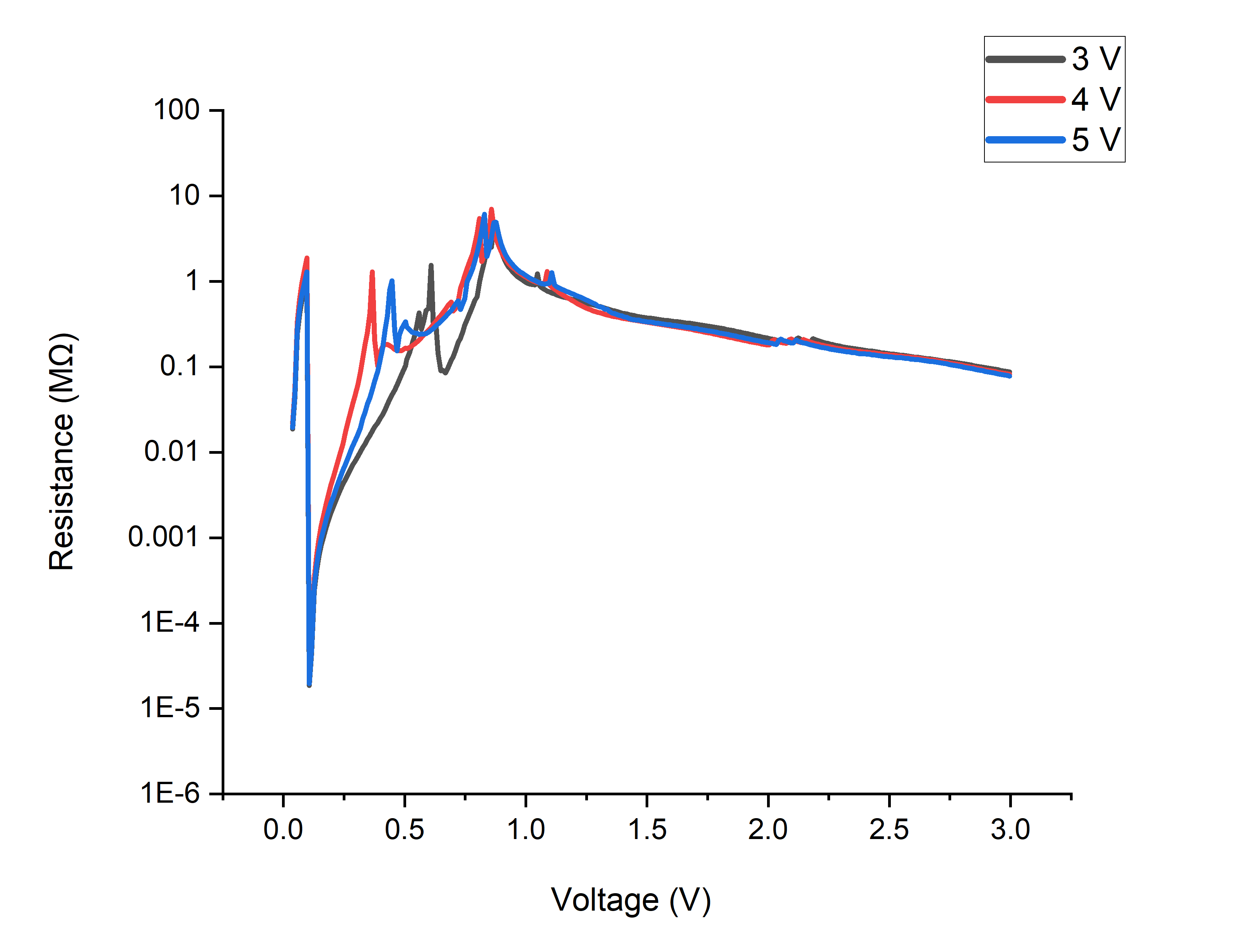}
\caption {Finding minimum voltage: Resistance of colloid with stimulation till 3~V, 4~V, and 5~V. There is a Fuchsian discontinuity that is cancelled by the modulus}
\label{fig04}
\end{figure}

In the second step, the minimum stimulation voltage for a resistance to drop were investigated. In this regard, 5~V, 4~V and 3~V DC stimulation were applied to the sample and then the resistance was measured. Figs.~\ref{fig04} indicates the the resistance change during the voltage rising. In all three graphs, the resistance starts to drop down at a voltage of about 0.7V and it decreases from $\sim$ \SI{7}{\mega\ohm} to $\sim$ \SI{78}{\kilo\ohm}.

For conductive particles filled polymer composites, the percolation hypothesis is commonly employed to describe the sharp change in electrical resistivity as a function of filler content~\cite{zhang2006temperature}. However, in our case, the filler content has been fixed. To distinguish the link between electrical resistivity and stimulation voltage and during of stimulation, from the filler concentration dependence of percolation, we refer to the stimulation dependence of percolation process as `percolation time'.
To discover the relation between duration of stimulation, critical voltage (a value of voltage at which the resistance starts to decrease) and resistance dropping point, the sample has been stimulated with several times and the resistance was measured in each step. All data are summarised in Tab.~\ref{tab01}.

\begin{table}[!tbp]
\scriptsize
\caption{Relation between the length of stimulation, critical voltage and resistance dropping point}
\begin{tabular}{ | m{1.8cm} | m{1.8cm}| m{1.8cm} | m{1.8cm} | m{1.8cm} | m{1.8cm}|} \hline
Each stimulus duration (ms) & Time of dropping R (s) & Stimulation length (s) & Highest R M($\Omega$) & Critical voltage (V) & Critical resistance (M$\Omega$) \\ [0.5ex] \hline\hline
1 & 1 & 10 & 1.21 & 1.55 & 1.21 \\ \hline
10 & 5 & 13 & 1.38 & 1.45 & 1.38 \\ \hline
100 & 8 & 22 & 1.37 & 1.40 & 1.37 \\ \hline
1000 & 22 & 111 & 1.76 & 0.95 & 1.76 \\ \hline
2000 & 29 & 211 & 2.83 & 0.65 & 2.83 \\ \hline
3000 & 49 & 312 & 3.96 & 0.75 & 3.96 \\ \hline
4000 & 63 & 412 & 3.18 & 0.75 & 3.18 \\ \hline
\end{tabular}
\label{tab01}
\end{table}

\begin{figure}[!tbp]
\centering
\includegraphics[width=\textwidth]{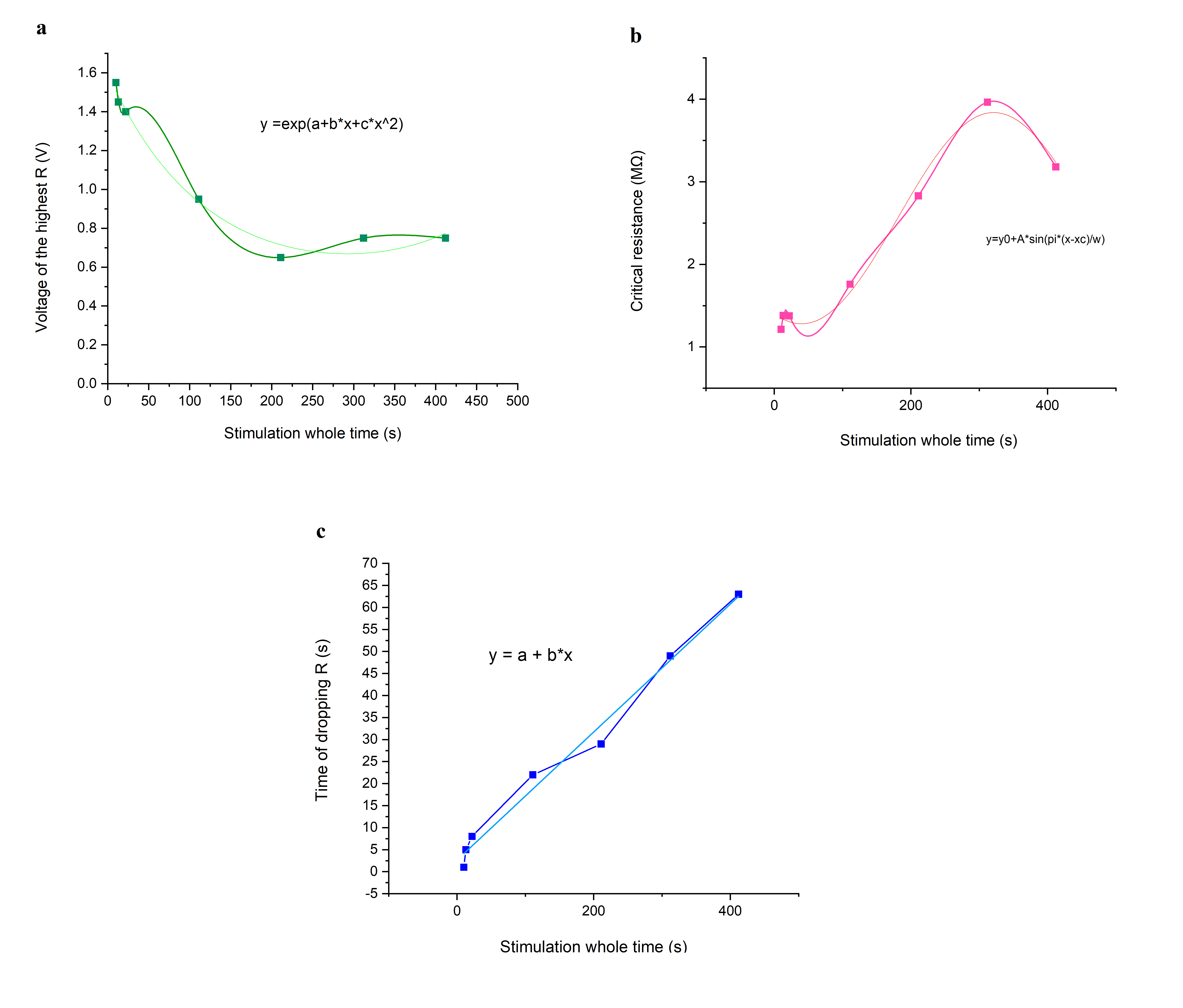}
\caption {a) Relation between critical voltage and length of stimulation, b) Relation between critical resistance and length of stimulation, c) Relation between length of stimulation and the time of dropping resistance}
\label{fig05}
\end{figure}

Fig.~(\ref{fig05}-a) shows the relation between critical voltage with length of stimulation. Critical voltage decrease with increase in stimulation length. The equation and parameters of fitting its curve are summarized in the table~\ref{tab02}.

\begin{table}[!tbp]
\scriptsize
\caption{The equation and parameters of fitting curve of critical voltage versus length of stimulation}
\begin{tabular}{ | m{3.6cm} | m{2cm} | m{2cm} | m{2cm} | m{1.7cm} | } \hline
Equation & parameter, $a$ & parameter, $b$ & parameter, c & R-Square (COD) \\ [0.5ex] \hline\hline
$y=exp(a+bx+cx^2)$ & 0.470 ± 0.028 & -0.006 ± 6.61E$^{-4}$ & 1.02E$^{-5}$ ± 1.71E$^{-6}$ & 0.98 \\ \hline
\end{tabular}
\label{tab02}
\end{table}

\begin{table}[!tbp]
\scriptsize
\caption{The equation and parameters of fitting curve of critical resistance versus length of stimulation}
\begin{tabular}{ | m{3.7cm} | m{1.5cm} | m{1.5cm} | m{1.5cm} | m{1.4cm} | m{1.3cm}| } \hline
Equation & parameter, y$-0$ & parameter, x$_c$ & parameter, W & parameter, A & R-Square (COD) \\ [0.5ex] \hline\hline
y=y$-0$+A sin(pi(x-x$_c$)/w) & 2.56 ± 0.068 & 180.51 ± 8.33 & 281.40 ± 18.14 & 1.28 ± 0.092 & 0.99 \\ \hline
\end{tabular}
\label{tab03}
\end{table}

Figure~\ref{fig05}b shows the relation between critical resistance with length of stimulation. Critical resistance rises with increase in stimulation length till reaches the amount of 3.96 M$\Omega$ and then with more enhancement in stimulation length, it falls again. The equation and parameters of fitting this curve are summarized in the Tab.~\ref{tab03}.

\begin{table}[!tbp]
\scriptsize
\caption{The equation and parameters of fitting curve of time of dropping resistance VS length of stimulation}
\begin{tabular}{ | m{3cm} | m{3cm} | m{3cm} | m{3cm} | } \hline
Equation & Intercept, a & slope, b & R-Square (COD) \\ [0.5ex] \hline\hline
y = a + bx & 2.64 ± 1.64 & 0.14 ± 0.008 & 0.99 \\ \hline
\end{tabular}
\label{tab04}
\end{table}

The filled conductive colloid are discovered to be thermodynamically non-equilibrium systems, in which the conductive network generation is time dependent, leading to the development of a concept known as dynamic percolation~\cite{zhang2006temperature}. In this research, the time of dropping resistance depends on the length of stimulation, linearly. This relation is shown in Fig.~(\ref{fig05}-c) and the fitting parameters are presented in Tab.~\ref{tab04}.

\begin{figure}[!tbp]
\includegraphics[width=\textwidth]{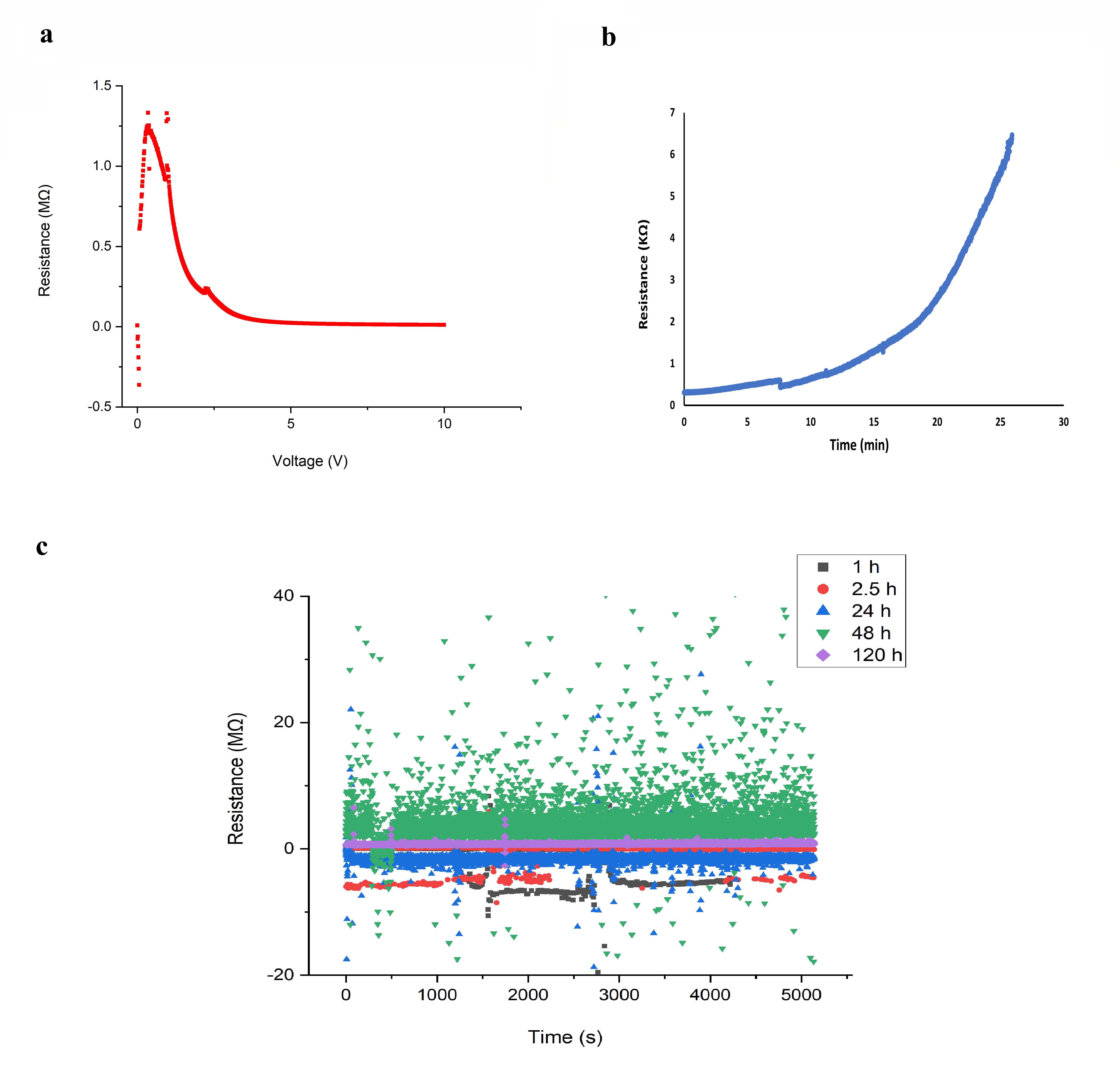}
\caption {Implementation of memory: a) Resistance of the colloid before stimulation, b) Resistance of the colloid with stimulation till 10V, c) Resistance of colloid after different hours}
\label{fig06}
\end{figure}

Electrical resistance before and after stimulation were compared to quantify both the quality and duration of the memory. To check the memorisation process, the resistance behaviour before and after stimulation, and find for how long does the stimulation effects last, some experiments were implemented. Firstly, the resistance of sample was measured without any stimulation other than the $400 \mu V$ applied by the measuring device, and then after applying DC stimulation, from \SIrange{0}{10}{\volt}, the resistance was measured again continuously (Fig.~\ref{fig06}a), to find the stability of the dropped resistance. Secondly, the stimulated sample was placed without any moving and any further stimulation. Then its resistance was measured just after stopping stimulation (Fig.~\ref{fig06}b), after 1h, 2.5h, 24h, 48h and 120h. All results are shown in the Fig.~(\ref{fig06}c).

Resistance was about 1.3 M$\Omega$ before applying any stimulation, during applying 10~V DC stimulation, resistance dropped down to under $\sim$ \SI{1}{\kilo\ohm} (Fig.~\ref{fig06}a). After stopping stimulation, the resistance started to rise from under \SI{1}{\kilo\ohm} to \SI{6.5}{\kilo\ohm} by spending 30 minutes (Fig.~\ref{fig06}b). In one hour, the resistance increased to \SI{96}{\kilo\ohm}, in 2.5 hours to \SI{128}{\kilo\ohm}, in 24 hours to \SI{370}{\kilo\ohm}, and in 120 hours to \SI{600}{\kilo\ohm} (Fig.~\ref{fig06}c).
As it can be seen, the resistance increases slowly during 120 hours after stopping stimulation indicating memory existence in our colloid.

\begin{figure}[!tbp]
\centering
\includegraphics[width=\textwidth]{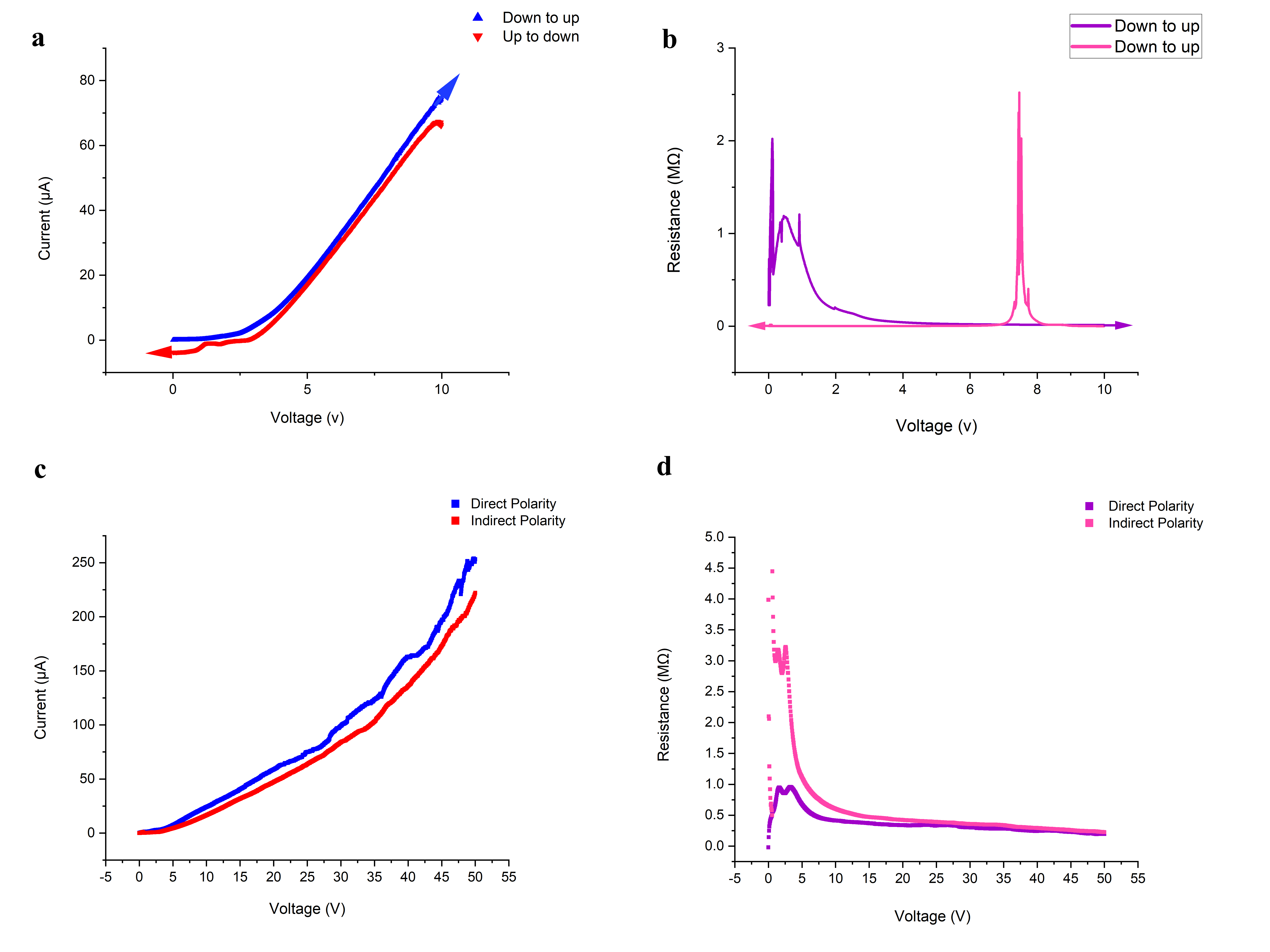}
\caption {a) Current diagram and b) Resistance diagram of colloid after DC stimulation from \SIrange{0}{10} {\volt} and then from  \SIrange{10}{0} {\volt}. c) Current diagram of colloid after 50V DC stimulation with direct and indirect polarity, d) Resistance diagram of colloid after 50V DC stimulation with direct and indirect polarity}
\label{fig07}
\end{figure}

In order to study the reverse cycle, the switching of the resistance back from low R to high R by reversing voltage were investigated. In the first stage, DC stimulation from \SIrange{0}{10}{\volt}, were applied with Keithley to the sample and then the resistance was measured. In the next stage, this process was repeated but in reverse cycle (applying DC stimulation from \SIrange{10}{0}{\volt}). Figures~\ref{fig07}a,b indicate the current diagram and the resistance diagram of this test, respectively.

In the next step, the polarity of electrodes was changed. The sample have been stimulated from \SIrange{0}{50}{\volt} with direct polarity, first. Then the cables on Keithley ports were swapped (without moving electrodes) and sample have been stimulated again from \SIrange{0}{50}{\volt} but with indirect polarity. The resulted current and resistance diagrams are shown in the Fig.~\ref{fig07}(c,d), respectively. The hysteresis loop of current and voltage shown in Fig.~\ref{fig07}c indicates history is more prominent in voltage and current approach zero.

\begin{figure}[!tbp]
\centering
\includegraphics[width=\textwidth]{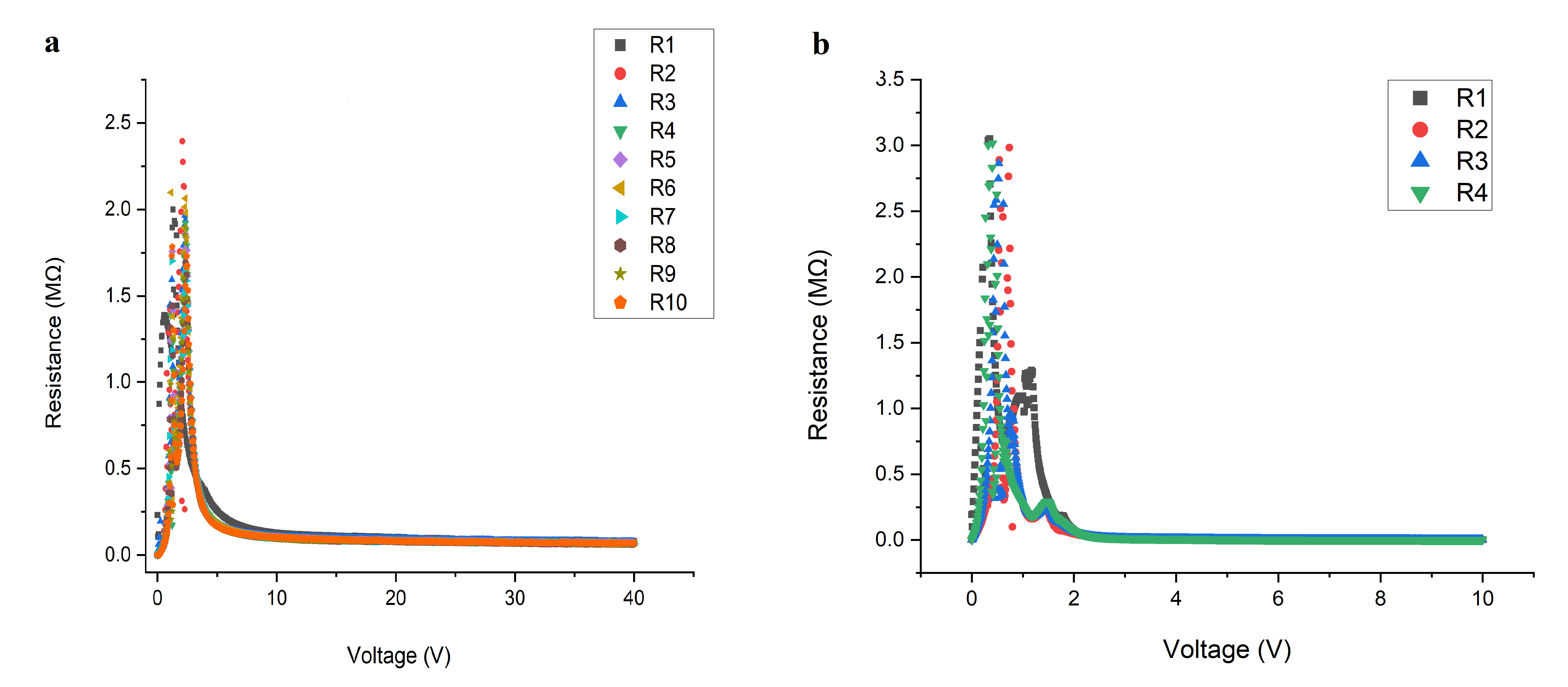}
\caption {a) Repeated measurement resistance of colloid with DC stimulation in the ambient atmosphere. b) Repeated measurement resistance of colloid with DC stimulation in the Nitrogen atmosphere and at $12^\circ C$}
\label{fig08}
\end{figure}

Finally, to check the effects of Brownian movements and oxygen interactions, we repeated the DC stimulation experiments under Nitrogen gas and at $12^\circ C$. This test was repeated four times. Fig.~\ref{fig08} shows the results of this experiment and compares them with the previous tests in an ambient atmosphere. No significant differences are observed between these two atmospheric conditions. Hence, we can conclude that Brownian movements and oxygen interactions do not have obvious effects on the learning and memorization process of the synthesized colloid.

\section{Conclusion}

We demonstrated that a colloid of zinc oxide in dimethyl sulfoxide is capable for formation of conductive pathways under application of direct current. The conductive pathways are volatile, they disintegrated with time. The phenomenon can be used to prototype memory devices in liquid phase robots and computers, and to implement synaptic devices in liquid phase neuromorphic circuits. Future directions of the research could be developing protocols for simulation formation of several conductive non-intersecting pathways in a spherical volume of a colloid. These pathways would play a role of information highways in liquid robots and computers.

\section{Acknowledgements}

The authors are grateful to David Patton for performing FESEM characterization and Paul Bowdler for his help to use the UV-Visible spectroscopy device.

This project has received funding from the European Union’s Horizon 2020 research and innovation programme FET OPEN ``Challenging current thinking” under grant agreement No 964388.

\end{document}